\documentclass{IEEEtran}
\usepackage{cite}
\usepackage{amsmath,amssymb,amsfonts}
\usepackage{graphicx}
\usepackage{textcomp,nicefrac}
\usepackage{siunitx}
\usepackage[switch]{lineno}
\usepackage{subfigure}
\def\BibTeX{{\rm B\kern-.05em{\sc i\kern-.025em b}\kern-.08em
T\kern-.1667em\lower.7ex\hbox{E}\kern-.125emX}}
\begin{document}
\title{Beam test of a baseline vertex detector prototype for CEPC}
\author{Shuqi Li, Tianya Wu, Xinhui Huang, Jia Zhou, Ziyue Yan, Wei Wang, Hao Zeng, Xuewei Jia, Yiming Hu, Xiaoxu Zhang, Zhijun Liang, Wei Wei, Ying Zhang, Xiaomin Wei, Lei Zhang, Ming Qi, Jun Hu, Jinyu Fu, Hongyu Zhang, Gang Li, Linghui Wu, Mingyi Dong, Xiaoting Li, Raimon Casanova, Liang Zhang, Jianing Dong, Jia Wang, Ran Zheng, Weiguo Lu, Sebastian Grinstein, Jo\~{a}o Guimar\~{a}es da Costa
\thanks{The research was supported and financed in large part by the National Key
Research and Development Program of China under Grant No. 2018YFA0404302
from the Ministry of Science and Technology. Additional support was provided by the
Youth Scientist Fund from the National Natural Science Foundation of China under
Grant No. 12205313.~\textit{(Shuqi Li and Tianya Wu contributed equally to this work.)}}
\thanks{Shuqi Li, Xinhui Huang, Jia Zhou, Ziyue Yan, Hao Zeng, Xuewei Jia, Mingyi Dong are with the Institute of High
Energy Physics, Chinese Academy of Sciences, Beijing 100049, China, and also with the University of Chinese Academy
of Sciences, Beijing 100049, China.}

\thanks{Tianya Wu is with the School of Information Engineering, Nanchang University, Nanchang 330031, China}

\thanks{Wei Wang, Zhijun Liang, Wei Wei, Ying Zhang, Jun Hu, Jinyu Fu, Hongyu Zhang, Gang Li, Linghui Wu, Weiguo Lu, Xiaoting Li, Jo\~{a}o Guimar\~{a}es da Costa are with the Institute of High
Energy Physics, Chinese Academy of Sciences, Beijing 100049, China (e-mail: liangzj@ihep.ac.cn, weiw@ihep.ac.cn, zhangying83@ihep.ac.cn).}
\thanks{Yiming Hu, Xiaoxu Zhang, Lei Zhang, Ming Qi are with the Department of Physics, Nanjing University, Nanjing 210093, China.}
\thanks{Xiaomin Wei, Jia Wang, Ran Zheng are with the Northwestern Polytechnical University, Xi'an, China.}
\thanks{Liang Zhang, Jianing Dong are with the Institute of Frontier and Interdisciplinary Science and Key Laboratory of Particle Physics and Particle Irradiation, Shandong University, Qingdao 266237, China.}
\thanks{Raimon Casanova, Sebastian Grinstein are with Institut  de F\'{i}sica d'Altes Energies (IFAE), Bellaterra (Barcelona), Spain. Sebastian Grinstein is also with the Catalan Institution for Research and Advanced Studies (ICREA), Barcelona, Spain.}

}

\maketitle

\begin{abstract}
The Circular Electron Positron Collider (CEPC) has been proposed to enable more thorough and precise measurements of the properties of Higgs, W, and Z bosons, as well as to search for new physics. In response to the stringent performance requirements of the vertex detector for the CEPC, a baseline vertex detector prototype was tested and characterized for the first time using a \SI{6}{\GeV} electron beam at DESY~\uppercase\expandafter{\romannumeral2} Test Beam Line 21. 
The baseline vertex detector prototype is designed with a cylindrical barrel structure that contains six double-sided detector modules (ladders). Each side of the ladder includes TaichuPix-3 sensors based on Monolithic Active Pixel Sensor (MAPS) technology, a flexible printed circuit, and a carbon fiber support structure. Additionally, the readout electronics and the Data Acquisition system were also examined during this beam test. The performance of the prototype was evaluated using an electron beam that passed through six ladders in a perpendicular direction. The offline data analysis indicates a spatial resolution of about \SI{5}{\um}, with detection efficiency exceeding \SI{99}{\%} and an impact parameter resolution of about \SI{5.1}{\um}. These promising results from this baseline vertex detector prototype mark a significant step toward realizing the optimal vertex detector for the CEPC.
\end{abstract}


\section{Introduction}
The CEPC is designed to operate at center-of-mass energies of \SI{91.2}{\GeV}, \SI{160}{\GeV}, and \SI{240}{\GeV}, serving as a Z-boson factory, reaching the threshold for WW pair production, and operating as a Higgs factory, respectively~\cite{CEPC-SPPC-detector}. 
The abundant production of $b \slash c-$quark jets during the CEPC operation highlights the critical role of flavor tagging in the design of the vertex detector. Effective flavor tagging requires accurate reconstruction of the vertex and the trajectory of charged tracks. Therefore, the physics goals of the CEPC are catalyzing the evolution of vertex detectors. The vertex detector for CEPC needs to achieve a single-point resolution better than \SI{3}{\um}, maintain a material budget below \SI{0.15}{\%} $X_0$ per layer, consume power below $\qty{50}{\mW\per\square\cm}$, and ensure a pixel sensor readout time shorter than \SI{10}{\us}~\cite{CEPC-SPPC-detector}. In striving to fulfill these requirements, a baseline vertex detector prototype has been developed and evaluated using an electron beam from DESY~\uppercase\expandafter{\romannumeral2}~\cite{DIENER:2019265}.

\begin{figure}[ht]
    \begin{center}
    \includegraphics[width=0.51\textwidth]{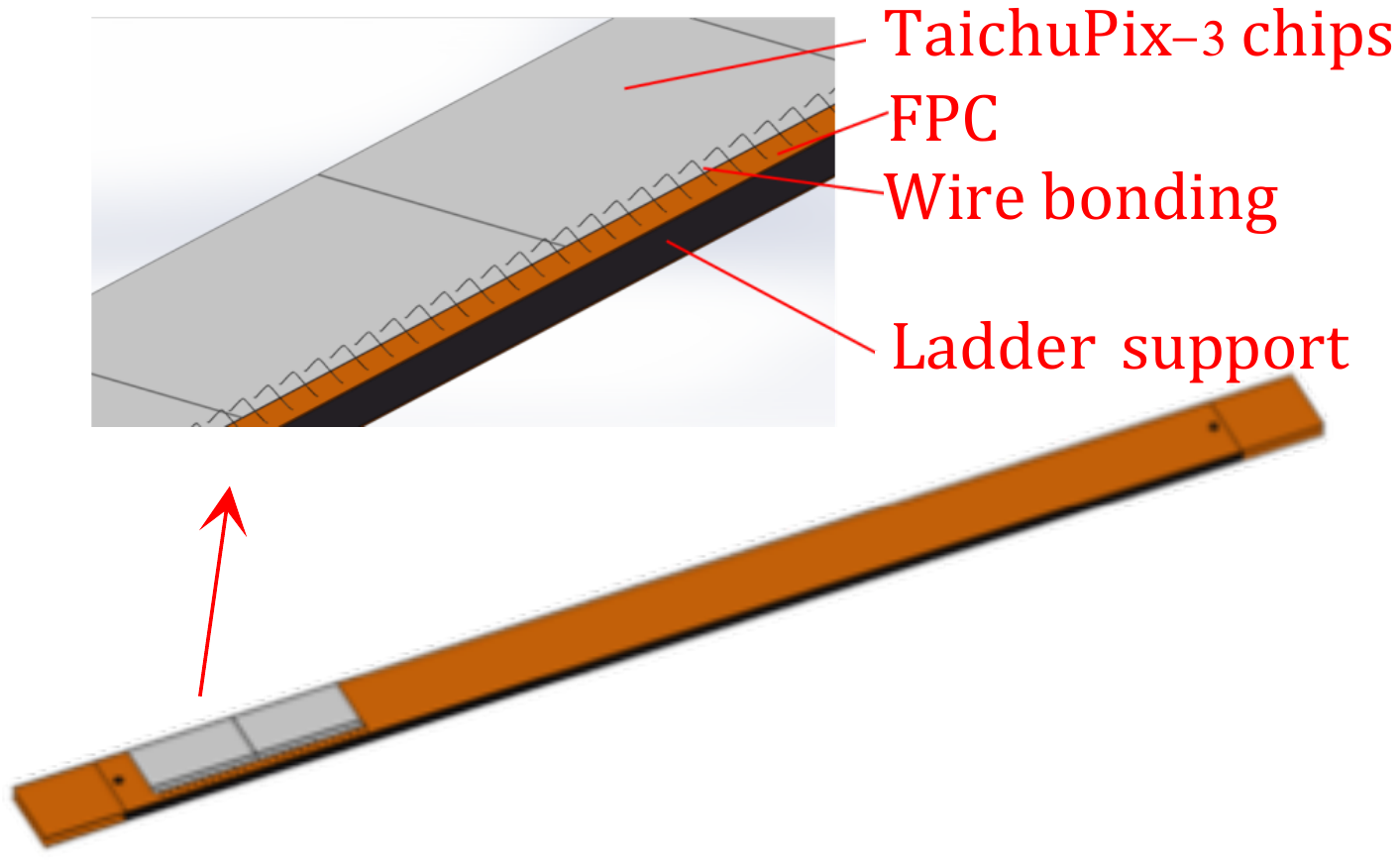}
\caption{The structure of the ladder~\cite{Fu2022MechanicalDO}.}
\label{fig:VTXD}
    \end{center}
\end{figure}

The baseline vertex detector prototype comprises three layers of concentric barrels positioned at radii of about \SI{18.1}{\mm}, \SI{36.6}{\mm}, and \SI{59.9}{\mm}.
The mechanical structure of the baseline vertex detector is fabricated to full scale following the design outlined in the CEPC Conceptual Design Report~\cite{CEPC-SPPC-detector}. The detector module, also known as the ladder, is a dual-sided structure that can place sensors on both sides.
The ladder consists of a pixel sensor chip, with up to ten on each side, a flexible printed circuit (FPC), and a support structure made of carbon fiber as depicted in Fig.~\ref{fig:VTXD}. Two sensors are wire-bonded onto the end of the FPC to cover the maximum area allowed by the collimator in DESY~\uppercase\expandafter{\romannumeral2}, which measures \qtyproduct{2.5x2.5}{\cm}. Controller, power, and data transfer are provided to the sensor by the FPC. The ladder has a thickness of approximately \SI{3.67}{\mm} with a material budget of about 1.6\%~$X_0$ and a length of about \SI{553}{\mm}. 
Six ladders are aligned and mounted along a certain azimuthal direction of the concentric barrels, as shown in Fig.~\ref{fig:prototype_axis}\subref{fig:prototype_side}.


%

Several prototype pixel sensors have been developed for the CEPC, such as the JadePix series with a rolling shutter readout architecture~\cite{Chen_2019,ZHOU2020164427}. In order to address the high hit density of the CEPC, a Monolithic Active Pixel Sensor (MAPS), named TaichuPix, has been developed with the goal of high readout speed and high spatial resolution. TaichuPix-1\&2 are used for technical validation and design optimization~\cite{Wu:2021mju,Zhang:2022rlo}. TaichuPix-3 is the first full-scale pixel chip designed for the vertex detector prototype. The TaichuPix-3 is manufactured using a \SI{180}{\nm} CMOS Imaging Sensor (CIS) technology. The dimensions of the chip are \qtyproduct{2.57x1.59}{\cm}. The chip contains 1024 columns $\times$ 512 rows with a pixel pitch of \SI{25}{\um}. 

Each pixel of the TaichuPix-3 chip integrates a sensing diode, an analog front-end, and a digital logical readout in each pixel. The analog front-end is designed based on the ALPIDE chip~\cite{AGLIERIRINELLA2017583}, which is developed for the upgrade of the ALICE Inner Tracking System (ITS)~\cite{Abelev:1625842}. In order to address the high hit rate of CEPC, the analog front-end of TaichuPix-3 has been specifically optimized to ensure a quicker response. In addition, the digital logical readout includes a hit storage register, logic for pixel mask, and test pulse configuration. The digital logical readout follows the FE-I3~\cite{PERIC2006178} designed for the ATLAS pixel detector~\cite{CERN-LHCC-97-016}, but it has been modified to adjust the pixel address generator and relocate the timestamp storage from within the pixel to the end of the column. This modification was necessary due to pixel size constraints. Furthermore, the double-column drain peripheral readout architecture of the TaichuPix-3 chip~\cite{Wei:2019wbr} employs an address encoder with a pull-up and pull-down network matrix~\cite{Wu:2021mju}.




Table~\ref{tab2} summarizes the design specifications for the TaichuPix-3 chip. Supposing that the average cluster size is 3, the maximum hit rate is calculated to be $36 \times 10^6/\text{cm}^2/\text{s}$ for the W operation mode in CEPC. A dead time of less than \SI{500}{\ns} is required to achieve a detection efficiency higher than 99\%. The power consumption is less than $\qty{200}{\mW\per\square\cm}$ when operating at a fast leading edge (\textless~\qty{200}{\ns}) of the analog front-end and a serializer interface of \qty{160}{\MHz}. The fake hit rate is less than~\SI{e{-12}}{\text{event}^{-1}\text{pixel}^{-1}} after masking noisy pixels~\cite{WU2024168945}.
The TaichuPix-3 is characterized by the utilization of two different processes, namely Process A and Process B. Process A is fabricated using the standard back-bias diode process and includes an extra deep N-layer mask compared to Process B, as detailed in Ref.\cite{W:2017technogy}. The performance of the pixel sensor chip has been verified with a \SI{4}{\GeV} electron beam at DESY~\uppercase\expandafter{\romannumeral2}. The intrinsic spatial resolution of Process A was found to be~\SI{4.8}{\um} and~\SI{4.5}{\um} for Process B, with a detection efficiency exceeding~\SI{99}{\%}, as reported in Ref.\cite{WU2024168945}. In total, 24 TaichuPix-3 sensors with a thickness of~\SI{150}{\um} were assembled into the baseline vertex detector prototype. 

\begin{table}[t]
\caption{Design specifications of the TaichuPix-3 chip.}
\centering
\label{tab2}
\begin{tabular}{l r}

\hline
Specification & Index \\
\hline

Pixel size&\qtyproduct{25x25}{\um}\\
Dimension&\qtyproduct{15.9x25.7}{\mm}\\
Techonology& CIS~\SI{180}{\nm} \\
Dead time& ~$<$~\SI{500}{\ns}\\
Power density&$<$~$\qty{200}{\mW\per\square\cm}$\\
Max. hit rate&$36 \times 10^6/\text{cm}^2/\text{s}$\\
Bunch spacing& Higgs: \SI{680}{\ns}; W: \SI{210}{\ns}; Z: \SI{25}{\ns}\\
\hline

\end{tabular}
\end{table}



 In order to evaluate the performance of the mechanical, electrical, and Data Acquisition system (DAQ) of the baseline vertex detector, a beam test was conducted in April 2023 at the DESY~\uppercase\expandafter{\romannumeral2} Test Beam Line 21 (TB~21)~\cite{DIENER:2019265}. The electron beam was directed through the six ladders installed on the prototype, generating precise reconstruction points in the TaichuPix-3 sensors. In this paper, the test beam setup is described in detail, and the characterization of the baseline vertex prototype obtained from the offline data analysis is reported and discussed.


\label{sec:introduction}


\section{Test beam setup}
\label{sec:TestBeamSetup}
\begin{figure}[ht]
    \begin{center}
    \includegraphics[width=0.52\textwidth]{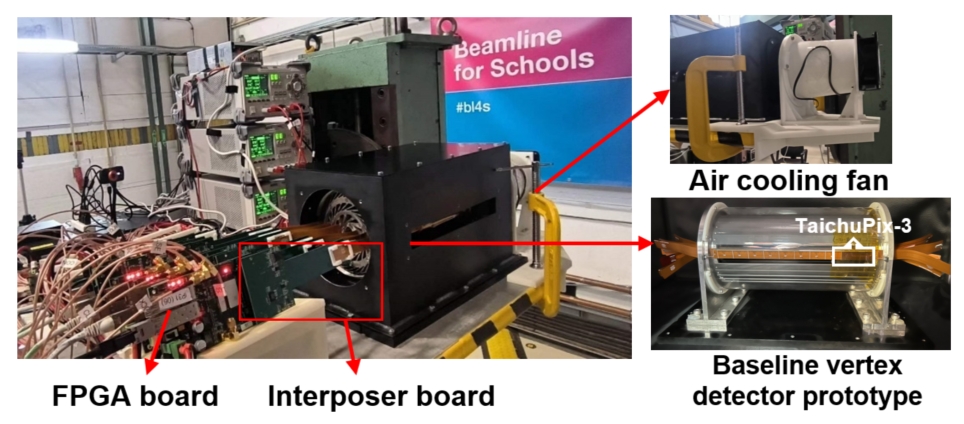}
    \caption{Baseline vertex detector prototype setup at DESY \uppercase\expandafter{\romannumeral2} TB21.}
    \label{fig:VTXDSETUP}
    \end{center}
\end{figure}

\begin{figure}[ht]
    \begin{center}
    \includegraphics[width=0.48\textwidth]{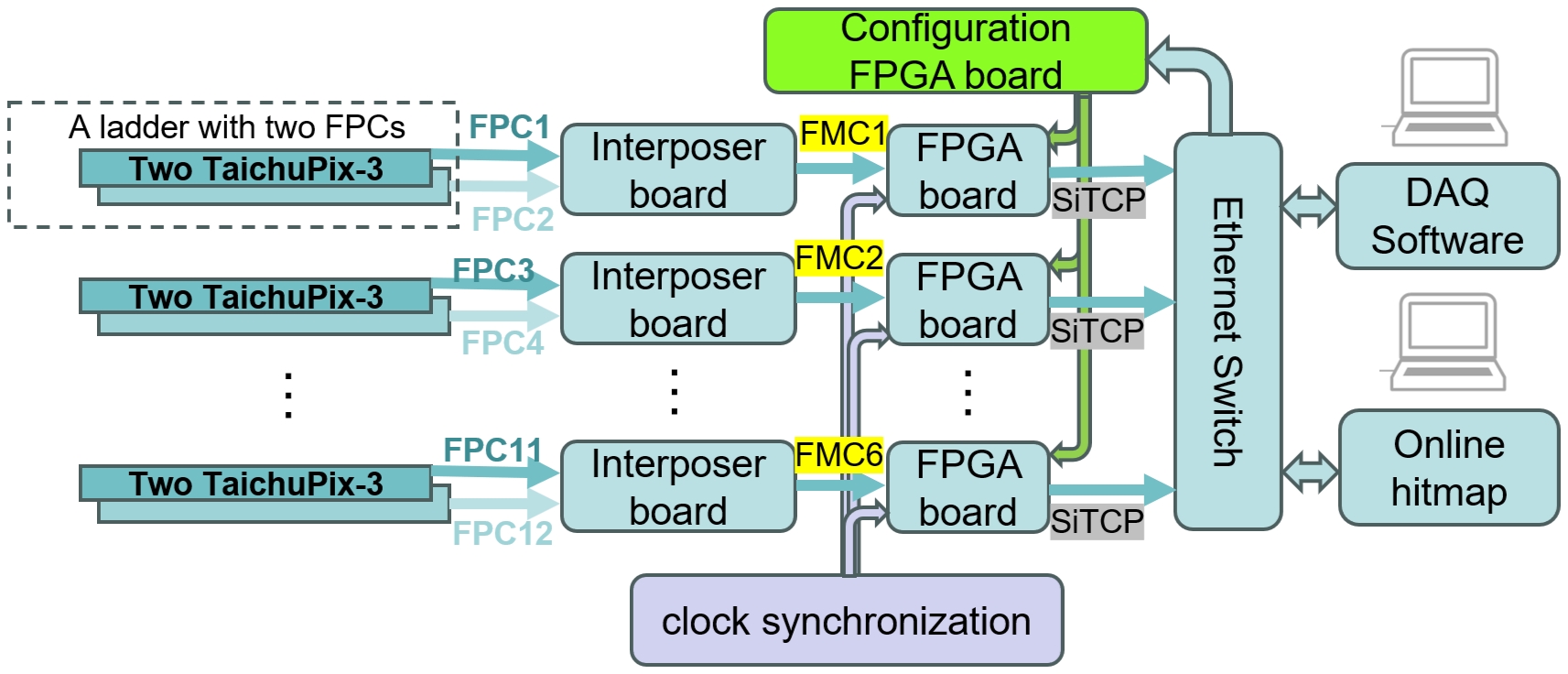}
    \caption{Scheme of the vertex detector setup at DESY \uppercase\expandafter{\romannumeral2} TB21. The FPC is the flexible printed circuit used to provide controller, power and data transfer between TaichuPix-3 chips and FPGA board. The FMC is the FPGA Mezzanine Card used to connect the Interposer board and FPGA board.}
    \label{fig:VTXDScheme}
    \end{center}
\end{figure}

The experimental setup is depicted in Fig.~\ref{fig:VTXDSETUP}. The prototype is placed within a black box, which includes an opening window on the side where the ladders are installed, enabling the beam to directly hit the ladders. The readout module of each ladder consists of an interposer board, an FPGA readout board, and a SiTCP protocol Ethernet port, as depicted in Fig.~\ref{fig:VTXDScheme}. The interposer board is used to transmit data from fired pixels and control signals between the ladder and the FPGA readout board. It also supplies DC voltages (\SI{1.8}{\V}) to the ladders. Each FPGA is controlled and synchronizes the clock through two synchronous ports: the clock controller port, and the global configuration port. 

The data package is transmitted through the Ethernet port to the switch and subsequently sent to the host computer. A dedicated DAQ system has been developed for the data collection. The DAQ software consists of two parts: data flow software and interactive software. The data flow software is the core of the entire system, responsible for data readout, transmission, online processing, and storage. The interactive software provides an intuitive operating interface for the experiment operator, implementing functions such as system status monitoring, message log processing, operation control, and online hitmap display. The DAQ employs a modular design and a module separation architecture, to reduce the mutual effect between different functional blocks and enhance the overall stability of the software. Furthermore, the highly modular design of the software facilitates straightforward upgrades and feature expansions to meet evolving experimental needs.

During the beam test, the TaichuPix-3 sensors were operated in a trigger-less mode and the back bias voltage was \SI{0}{\V}. The readout system operated reliably throughout all production runs and the recorded maximum data rate was about $18$ MB$\cdot$\SI{}{\s}$^{-1}$. An electric fan was utilized to cool the prototype as depicted in Fig.~\ref{fig:VTXDSETUP}, effectively reducing the temperature of the outermost layer from \SI{40}{\degreeCelsius} to \SI{28}{\degreeCelsius}, as measured with an infrared camera.

\begin{figure}[ht]
    \begin{center}
    \subfigure[]{\includegraphics[width=0.42\textwidth]{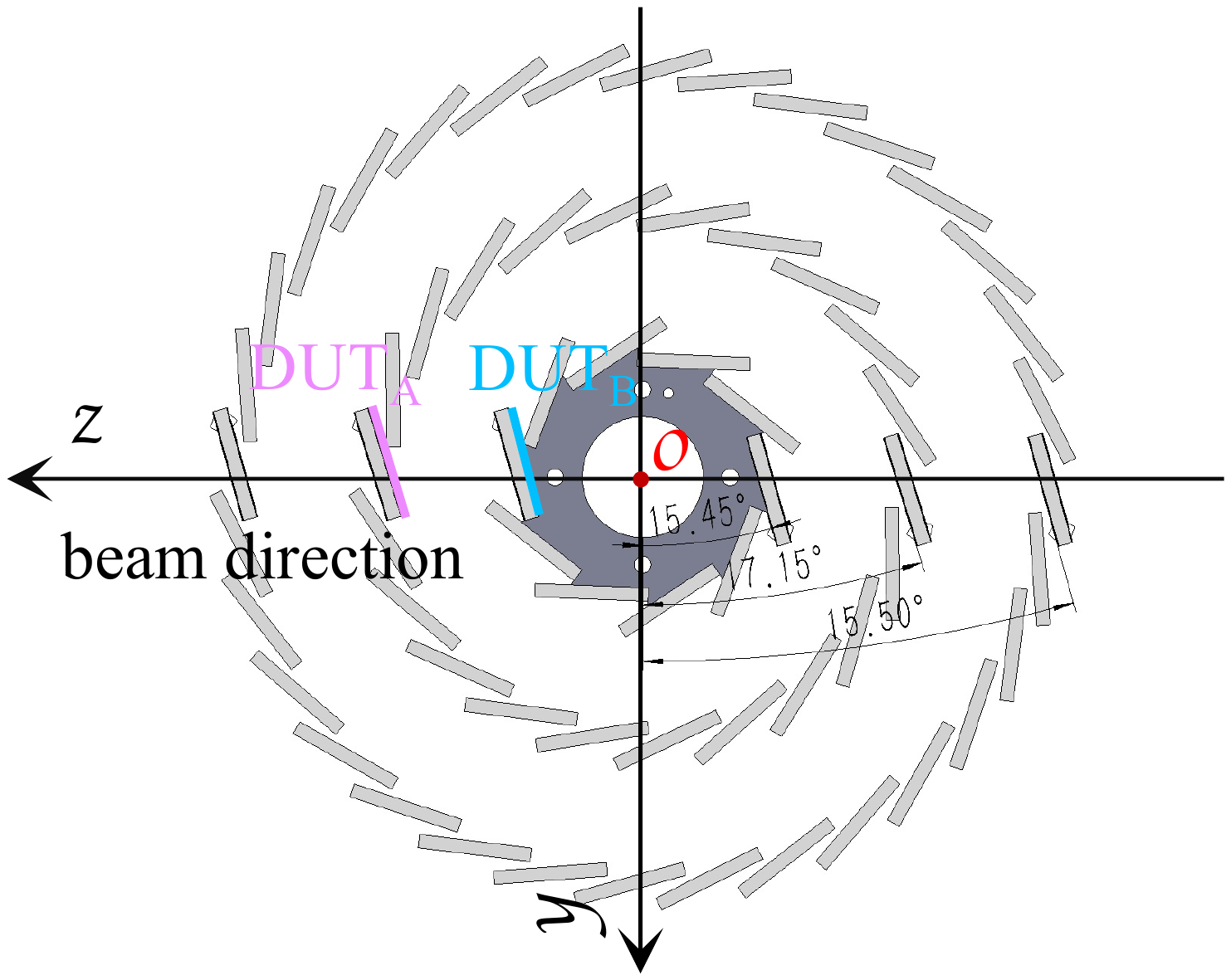}\label{fig:prototype_front} }
    \subfigure[]{\includegraphics[width=0.38\textwidth]{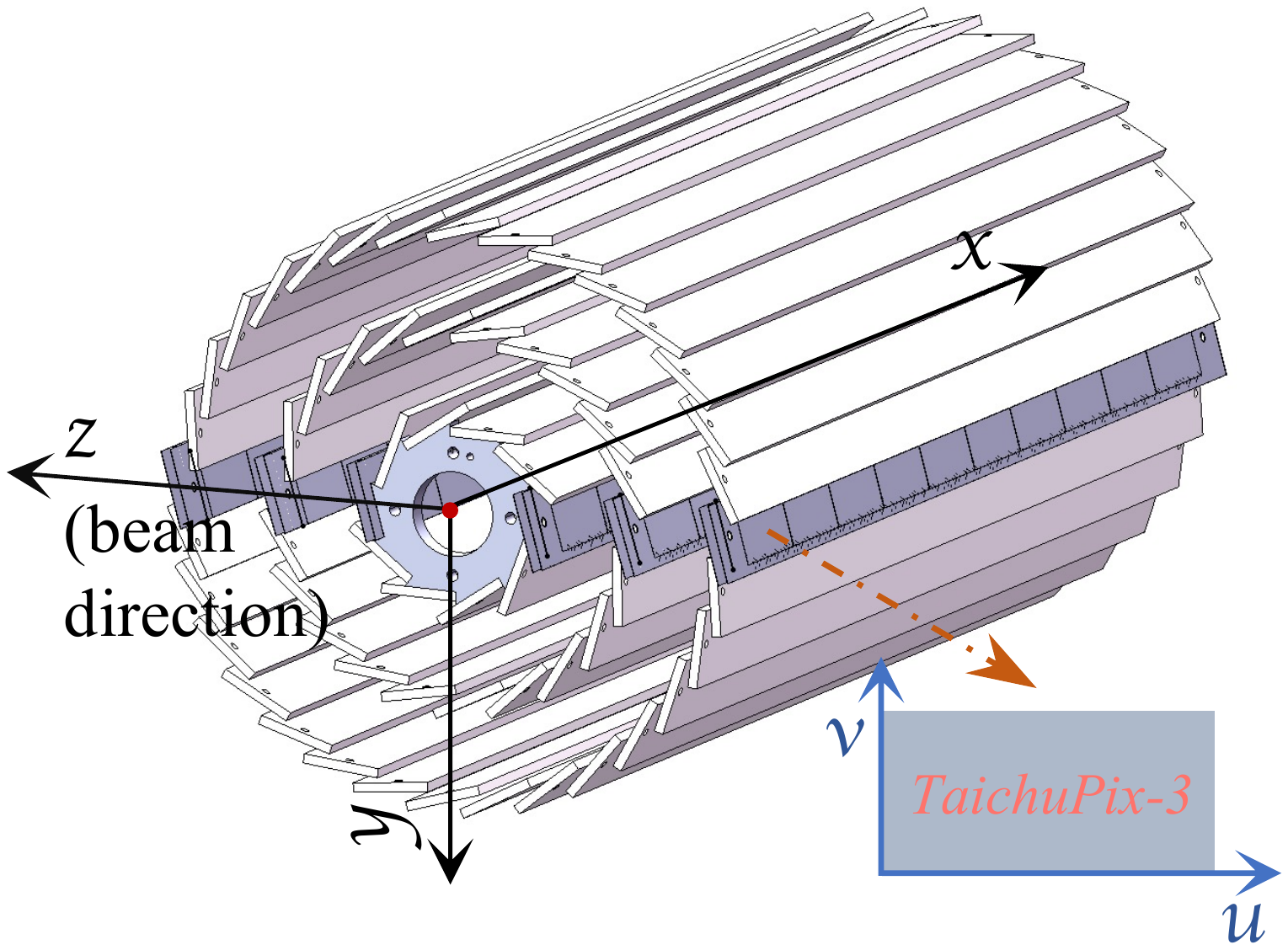}\label{fig:prototype_side} }
    \caption{\subref{fig:prototype_front} Position of DUT$_{\rm A}$ and DUT$_{\rm B}$ is highlighted with purple and blue color, respectively. The global coordinate system has the $z$-axis aligned with the electron beam direction. Along the beam direction, the process types of the TaichuPix-3 chips on the six ladders are B, A, B, B, A, B, respectively. ~\subref{fig:prototype_side}
    Definition of the local coordinate system on each TaichuPix-3 chip, where the $u$-direction runs along the row direction of the chip, and the $v$-direction runs along the column direction of the chip.}
    \label{fig:prototype_axis}
    \end{center}
\end{figure}

Fig.~\ref{fig:prototype_axis} illustrates the definition of global and local coordinates used in the offline analysis. The inclination angles of the six ladders, in the direction of the beam, are \SI{15.50}{\degree}, \SI{17.15}{\degree}, \SI{15.45}{\degree}, \SI{15.45}{\degree}, \SI{17.15}{\degree}, and \SI{15.50}{\degree}, respectively. The analysis of the offline data is based on TaichuPix-3 sensors based on Process A and Process B, which are positioned as shown in Fig.~\ref{fig:prototype_front}, and labeled as DUT$_{\rm A}$ and DUT$_{\rm B}$, respectively. When one Detector Under Test (DUT) is under study, the other planes are used to determine the reference tracks. 

\begin{figure}[ht]
    \begin{center}
    \includegraphics[width=0.51\textwidth]{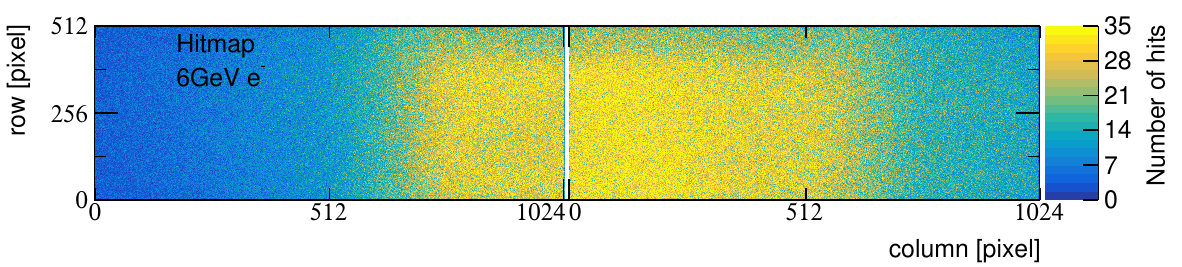}
\caption{The hitmap obtained with the 6 GeV electron beam.}
\label{fig:ladder_hitmap}
    \end{center}
\end{figure}

An example of hitmap is depicted in Fig.~\ref{fig:ladder_hitmap}, demonstrating the proper functioning of the entire detection system.

\section{Offline analysis and results}
\label{sec:Offline analysis and results}
The offline analysis procedure consists of several steps, including decoding raw data, clustering, track finding and reconstruction, and alignment of the detector geometry. Specifically, clustering is the process of grouping adjacent pixels with the collected charge above the set threshold $\xi$, and the center of the cluster is calculated using the Center of Gravity (CoG) method with a weight of 1, since the pixel is binary readout. The tracks are reconstructed using the General Broken Line (GBL) package~\cite{KLEINWORT2012107}, which accounts for multiple scattering. The geometry of the prototype is aligned using the Millepede algorithm~\cite{BLOBEL20065}, with the alignment parameters consisting of three translations and three rotations for each sensor. These alignment parameters are determined by minimizing the track-to-hit residual calculated by the track model. The masked noisy pixels are not excluded from the offline analysis, as the number of masked pixels is around 0.02\% of all pixels, which is small enough to neglect the effects on the results. 


\subsection{Cluster Size}

\begin{figure}[ht]
    \begin{center}
    \subfigure[]{\includegraphics[width=0.23\textwidth]{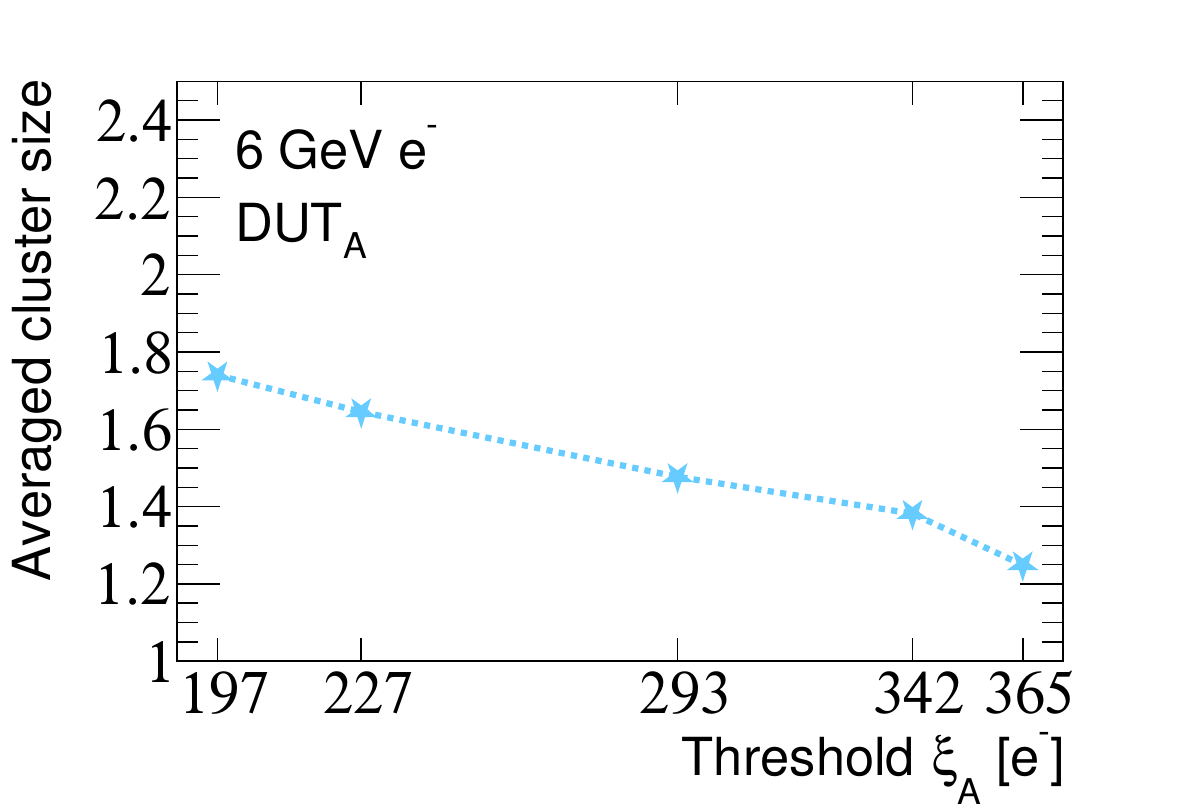}\label{fig:standard_thr} }
    \subfigure[]{\includegraphics[width=0.23\textwidth]{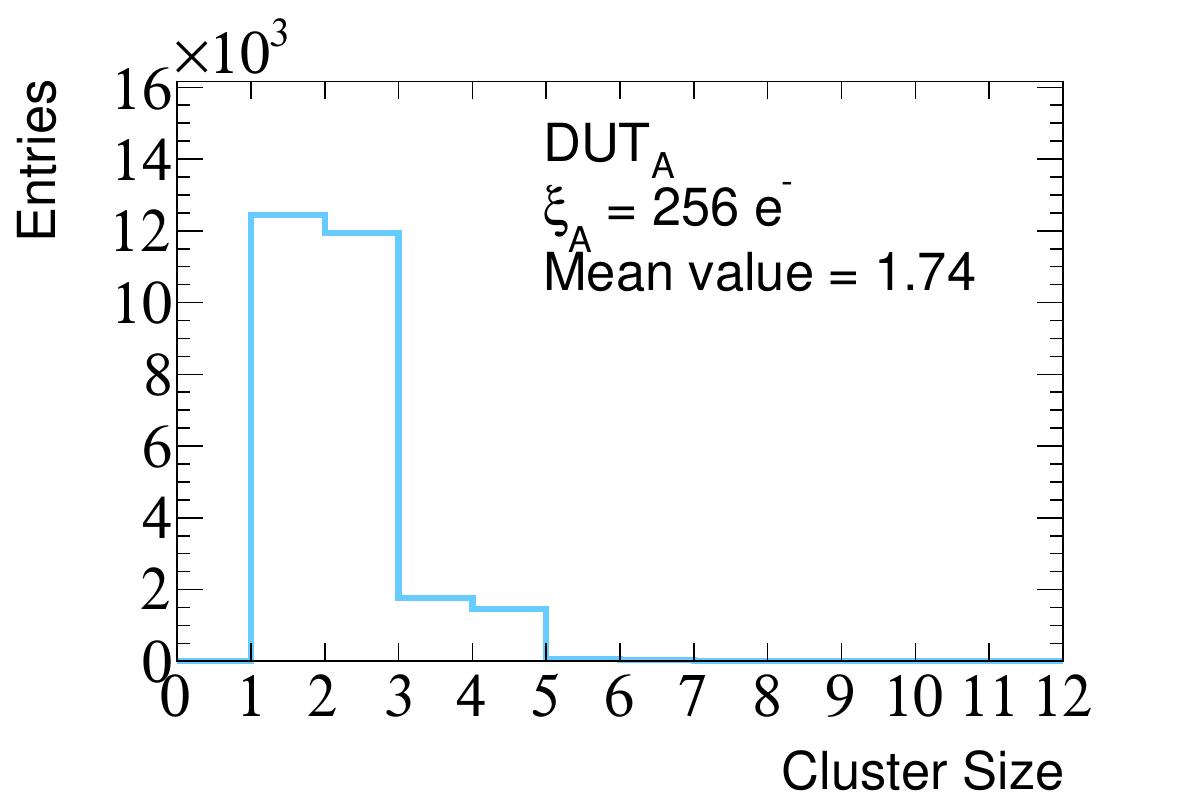}\label{fig:standard_clusdis} } \\
    \subfigure[]{\includegraphics[width=0.23\textwidth]{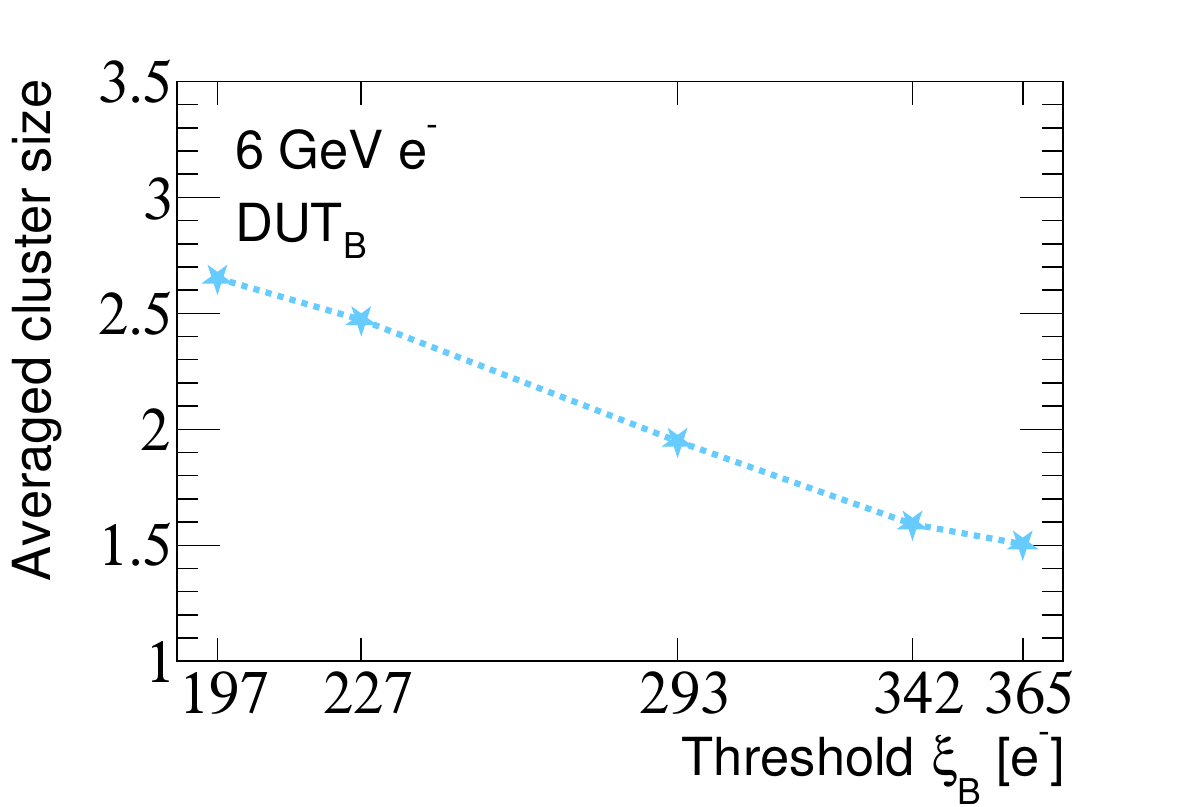}\label{fig:modify_thr} }
    \subfigure[]{\includegraphics[width=0.23\textwidth]{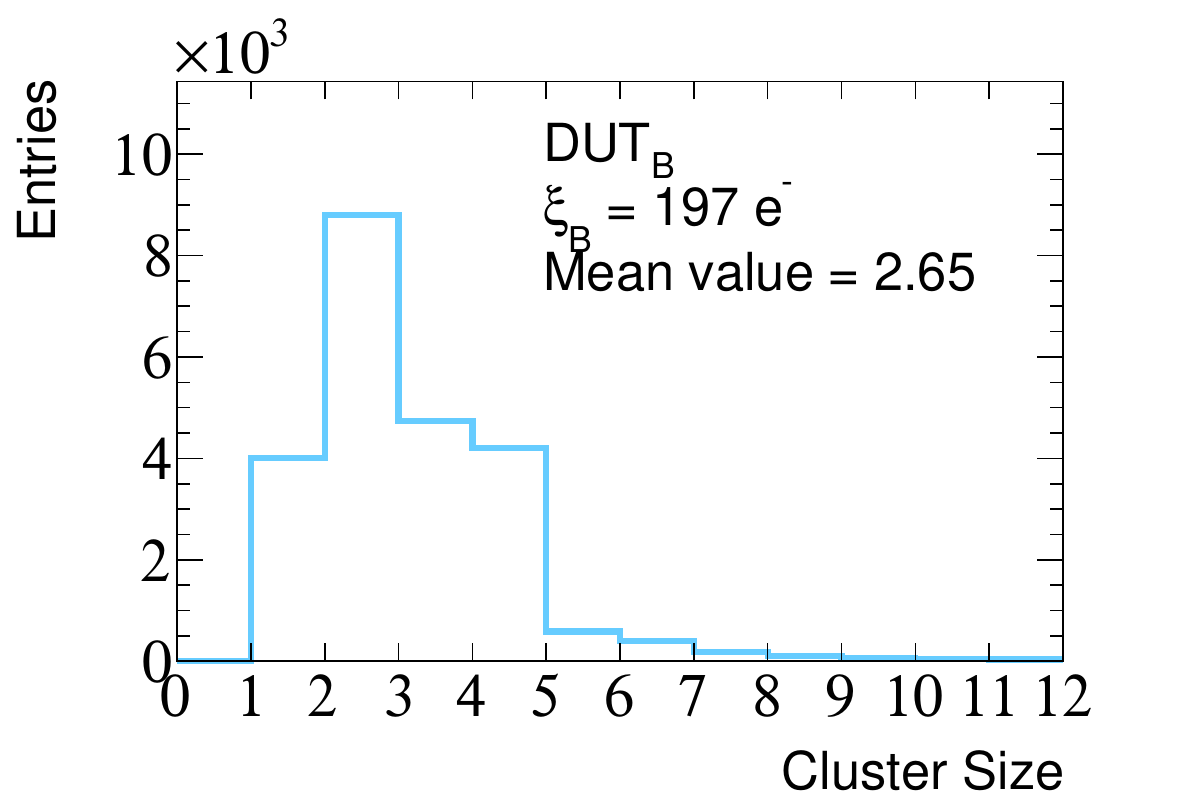}\label{fig:modify_clusdis} }
    \caption{Variation of cluster size with threshold for DUT$_{\rm A}$~\subref{fig:standard_thr} and DUT$_{\rm B}$~\subref{fig:modify_thr}. The cluster size distribution for DUT$_{\rm A}$~\subref{fig:standard_clusdis} and DUT$_{\rm B}$~\subref{fig:modify_clusdis} at the lowest threshold.
}
    \label{fig:ClusterSize_threshold}
    \end{center}
\end{figure}

The cluster size is the number of neighboring fired pixels with signals above a certain threshold $\xi$. A higher threshold leads to a reduction in the number of fired pixels, consequently weakening the charge-sharing effect and resulting in a deterioration in spatial resolution. Fig.~\ref{fig:ClusterSize_threshold} shows the average cluster size for DUT$_{\rm A}$ and DUT$_{\rm B}$, which is drawn as a function of threshold $\xi_{\rm A}$ and $\xi_{\rm B}$, respectively. The cluster size decreases as the threshold increases. At the minimum threshold, the averaged cluster size for DUT$_{\rm A}$ and DUT$_{\rm B}$ is 1.74 pixels and 2.65 pixels, respectively. The smaller mean cluster size in DUT$_{\rm A}$ relative to DUT$_{\rm B}$ indicates a reduced charge-sharing effect in  DUT$_{\rm A}$. This is because an additional low dose deep n-type is implemented in the epitaxial layer in DUT$_{\rm A}$, which will result in a larger depletion region than in DUT$_{\rm B}$, as discussed in Ref.~\cite{W:2017technogy}. A larger depletion region reduces the probability of electron-hole pairs diffusing to other pixels, thus weakening the charge sharing effect.

\subsection{Spatial resolution}
\begin{figure}[ht]
    \begin{center}
    \subfigure[]{\includegraphics[width=0.23\textwidth]{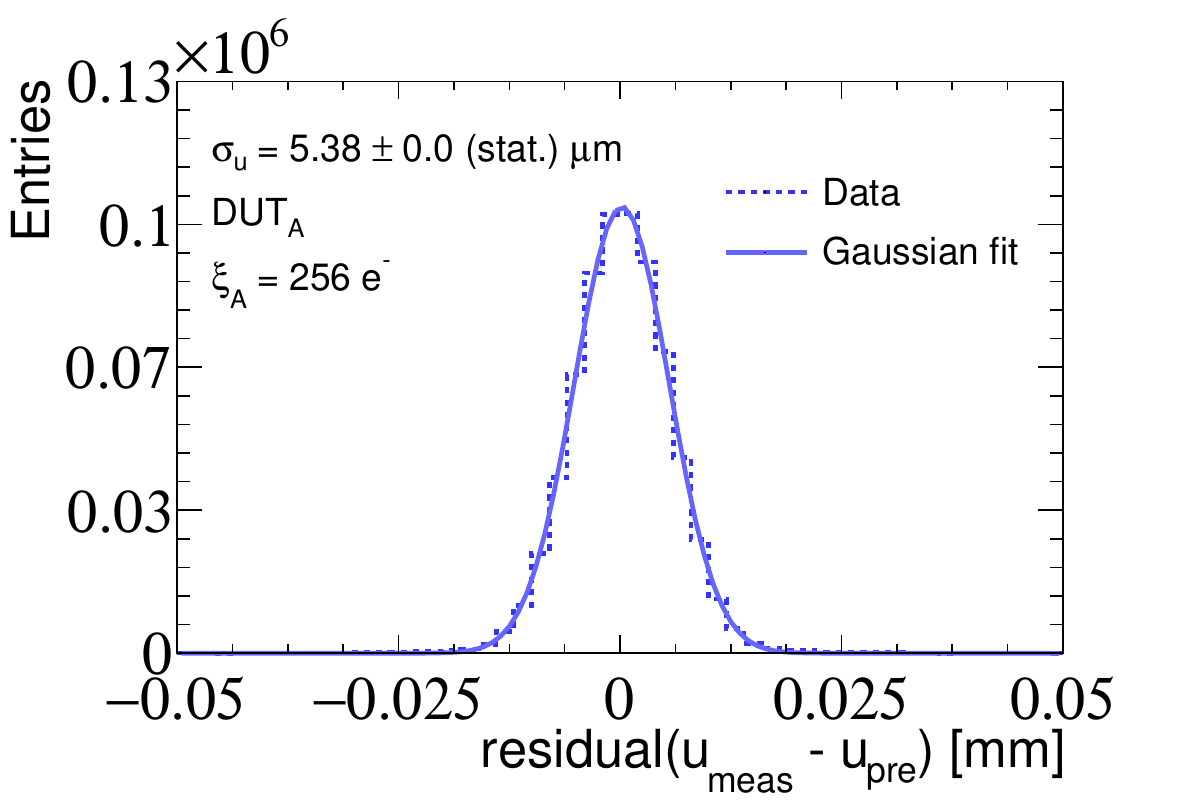}\label{fig:x_res_distribution} }
    \subfigure[]{\includegraphics[width=0.23\textwidth]{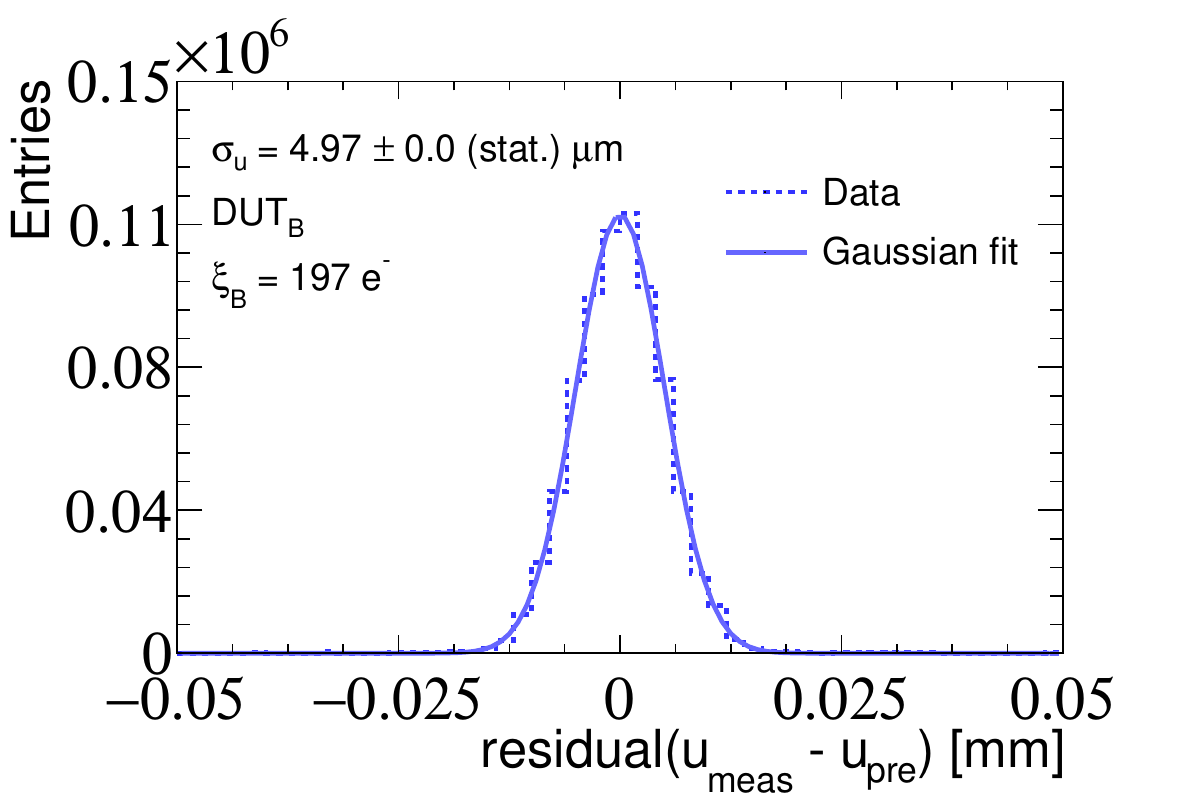}\label{fig:y_res_distribution} }
    \caption{\subref{fig:x_res_distribution} The unbiased residual distribution in DUT$_{\rm A}$ \subref{fig:x_res_distribution} and DUT$_{\rm B}$ \subref{fig:y_res_distribution} along $u$-direction.}
    \label{fig:residual_distribution}
    \end{center}
\end{figure}

\begin{figure}[ht]
    \begin{center}
    \subfigure[]{\includegraphics[width=0.23\textwidth]{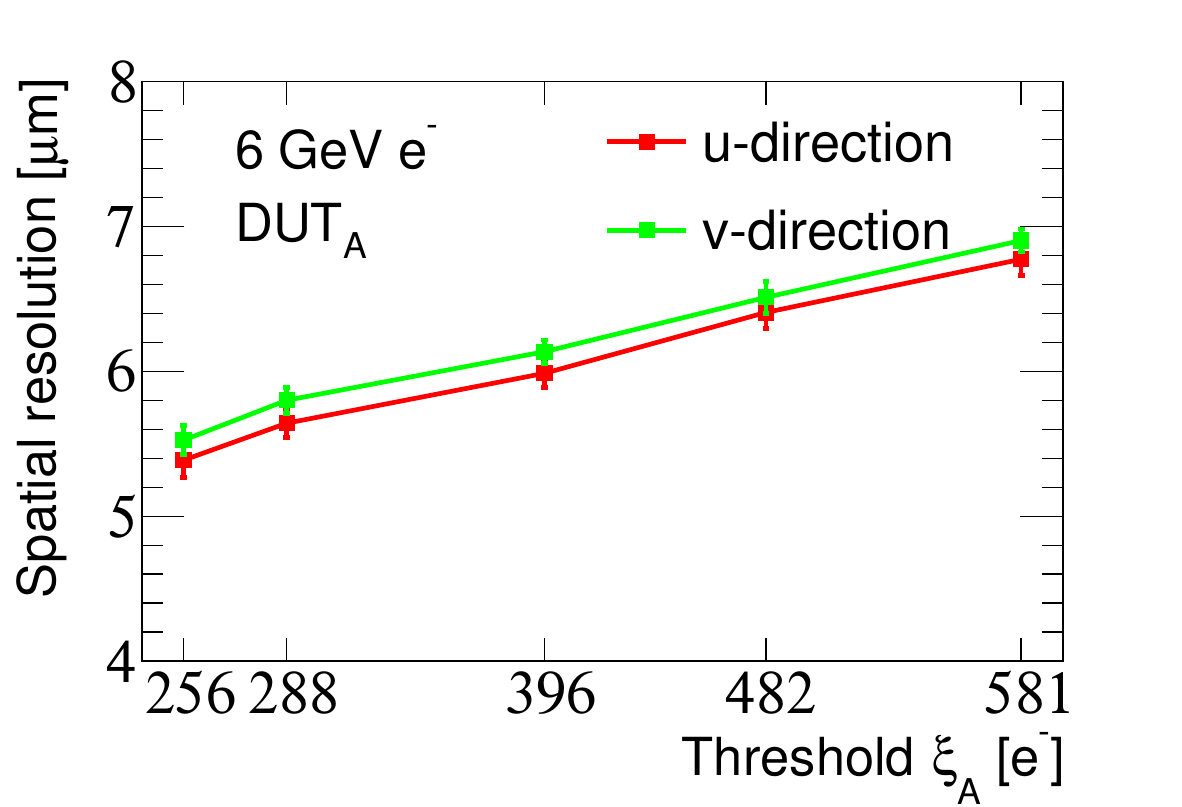}\label{fig:mod_res_thr} }
    \subfigure[]{\includegraphics[width=0.23\textwidth]{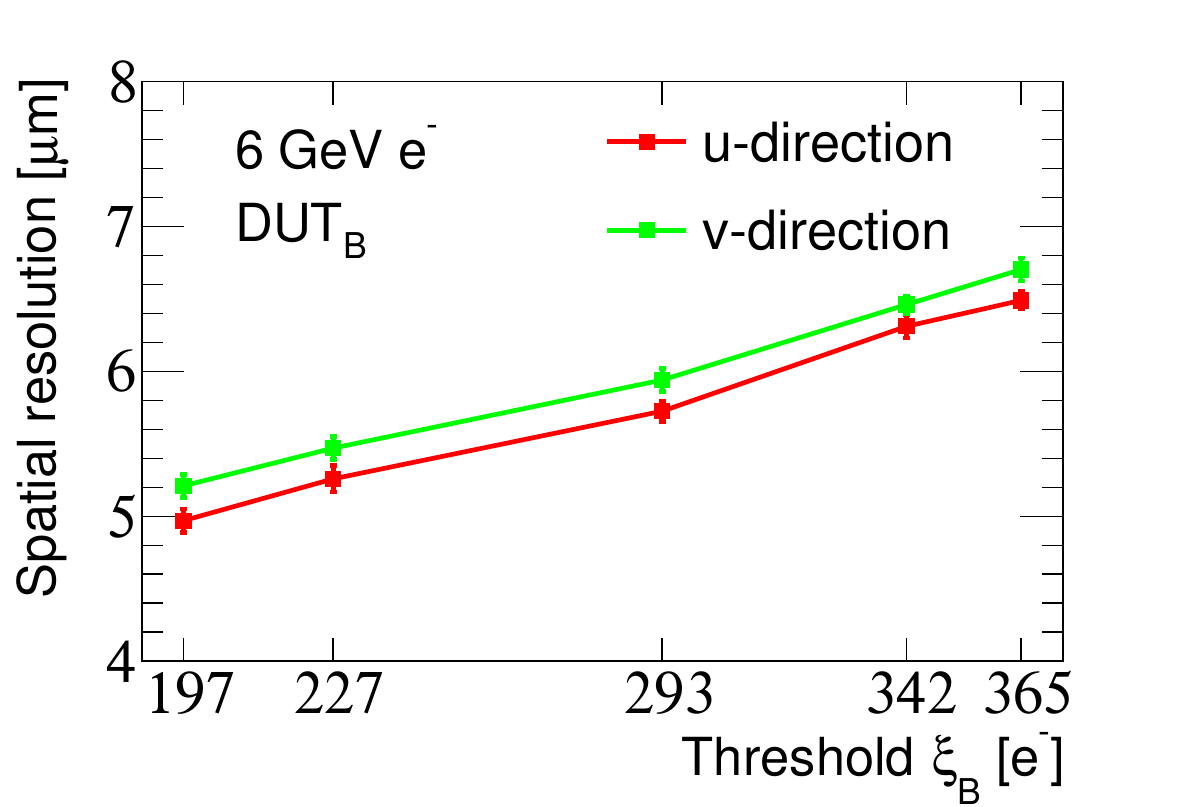}\label{fig:std_res_thr} }
    \caption{The variation of spatial resolution with threshold for DUT$_{\rm A}$ and DUT$_{\rm B}$. The error bars represent the systematic uncertainty from the beam energy spread (5~\%)~\cite{KLEINWORT2012107} and the accuracy of the scattering angle predicted by Highland formula (11~\%)~\cite{HIGHLAND1975497}. The statistical error is small enough to be negligible.}
    \label{fig:res_thr}
    \end{center}
\end{figure}

The spatial resolution is derived from an unbiased distribution of tracking residual obtained using the GBL track-fitting algorithm, which excludes the DUT~\cite{KLEINWORT2012107}. The scattering angle is predicted using the Highland formula~\cite{HIGHLAND1975497}. After alignment, the difference between the predicted and measured hit positions on the DUT is shown in Fig.~\ref{fig:residual_distribution}. The standard deviation for DUT$_{\rm A}$ and DUT$_{\rm B}$ at minimum threshold is approximately \SI{5.4}{\um} and \SI{5.0}{\um}, respectively.  Additionally, as depicted in Fig.~\ref{fig:res_thr}, the spatial resolution of both DUTs deteriorates as the threshold increases, and due to reduced charge-sharing effects on DUT$_{\rm A}$, it exhibits poorer resolution compared to DUT$_{\rm B}$. At the lowest threshold setting, the best spatial resolution achieved is $5.38 \pm 0.12$ (syst.) \SI{}{\um} in the $u$-direction and $5.52 \pm 0.10$ (syst.) \SI{}{\um} in the $v$-direction for DUT$_{\rm A}$, and $4.97 \pm 0.08$ (syst.) \SI{}{\um} in the $u$-direction and $5.21 \pm 0.08$ (syst.) \SI{}{\um} in the $v$-direction for DUT$_{\rm B}$.

\begin{figure}[ht]
    \begin{center}
    \includegraphics[width=0.35\textwidth]{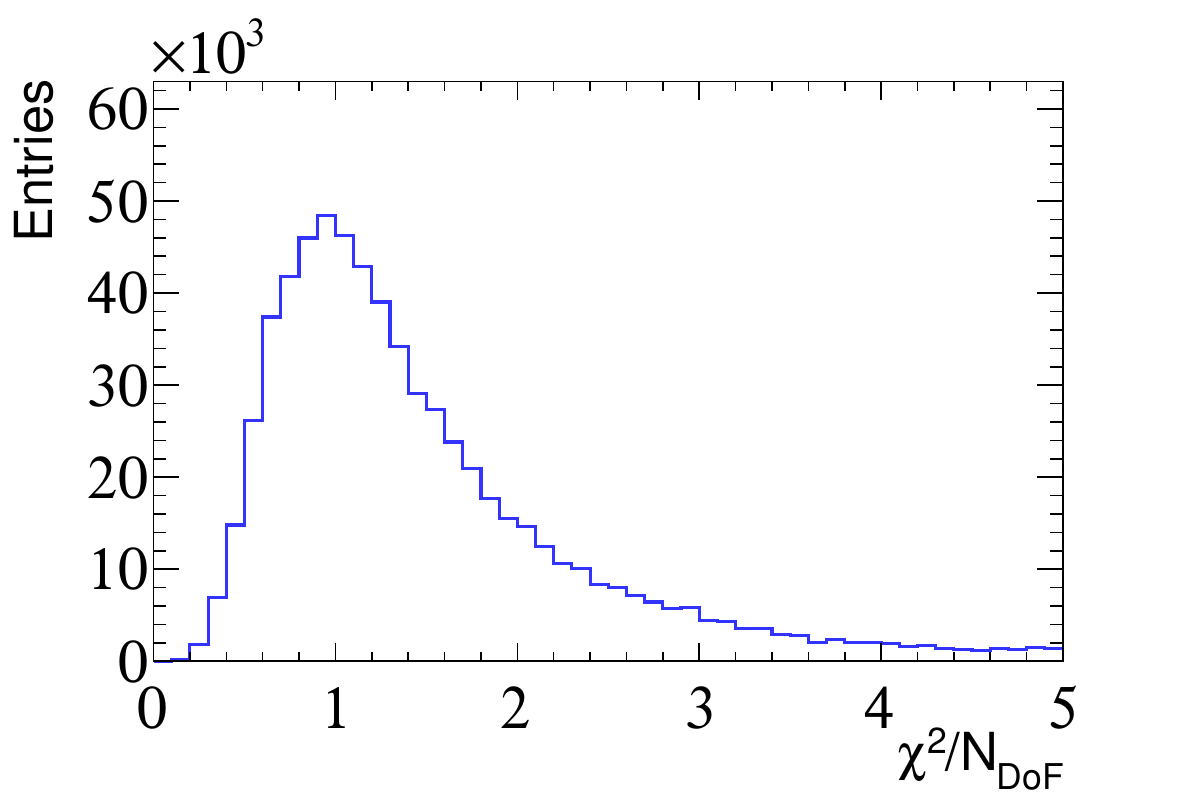}
\caption{Distribution of the $\chi^2$ per degree of freedom.}
\label{fig:chi2}
    \end{center}
\end{figure}

\subsection{Detection efficiency}
The detection efficiency $\varepsilon$ is defined as the ratio of the number of tracks that can match the measured points on the DUT $\rm (N_{matched}^{tracks})$ to the total number of tracks $\rm (N_{all}^{tracks})$. For matched tracks, the extrapolated hit position on the DUT is required to be less than
the specified distance $d$ from the measured hit position. In this analysis, $d$ is set to \SI{100}{\um} to exclude poorly reconstructed tracks. The detection efficiency can be expressed as follows:

\begin{equation}
\label{eq7}
\varepsilon = \frac{N_{|x_{meas}, y_{meas}-x_{pre}, y_{pre}| < d}^{matched\;tracks }}{N_{all}^{tracks}}
\end{equation}

\begin{figure}[ht]
    \begin{center}
    \subfigure[]{\includegraphics[width=0.23\textwidth]{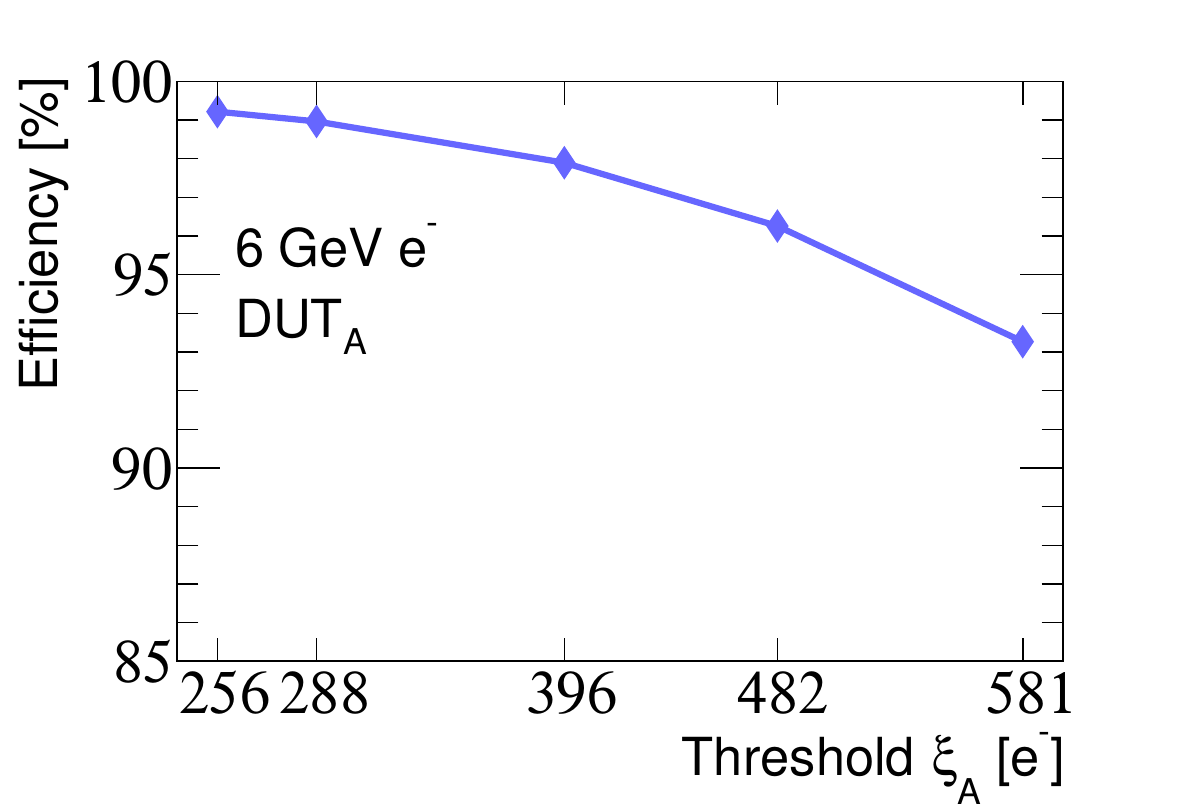}\label{fig:eff_A} }
    \subfigure[]{\includegraphics[width=0.23\textwidth]{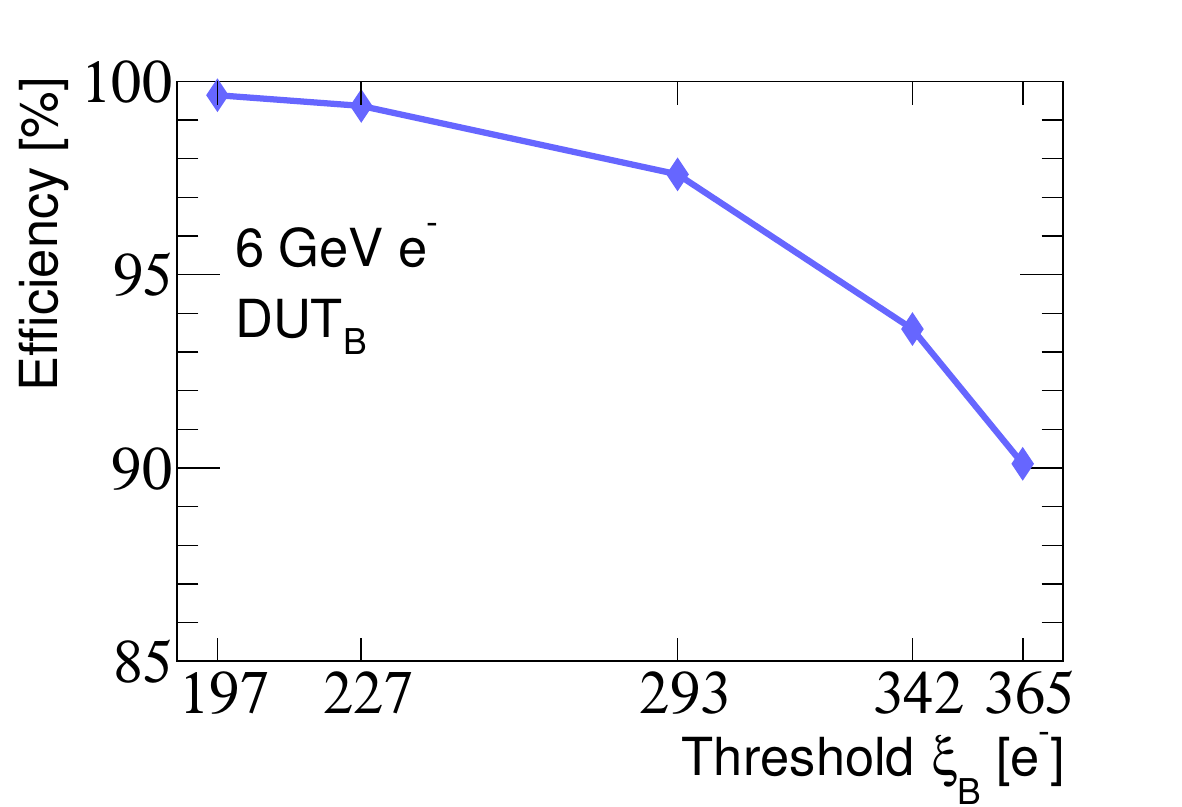}\label{fig:eff_B}}
    \caption{The detection efficiency of DUT$_{\rm A}$ and DUT$_{\rm B}$ as a function of the threshold $\xi_{\rm A}$ and $\xi_{\rm B}$.}
    \label{fig:Eff_Threshold}
    \end{center}
\end{figure}

As shown in Fig.~\ref{fig:Eff_Threshold}, the efficiencies of DUT$_{\rm A}$ and DUT$_{\rm B}$ exhibit a decreasing trend as the threshold increases. The best detection efficiency is \SI{99.3}{\%} and \SI{99.6}{\%} for DUT$_{\rm A}$ and DUT$_{\rm B}$, respectively. 

\subsection{Impact parameters}

\begin{figure}[ht]
    \begin{center}
    \subfigure[]{\includegraphics[width=0.35\textwidth]{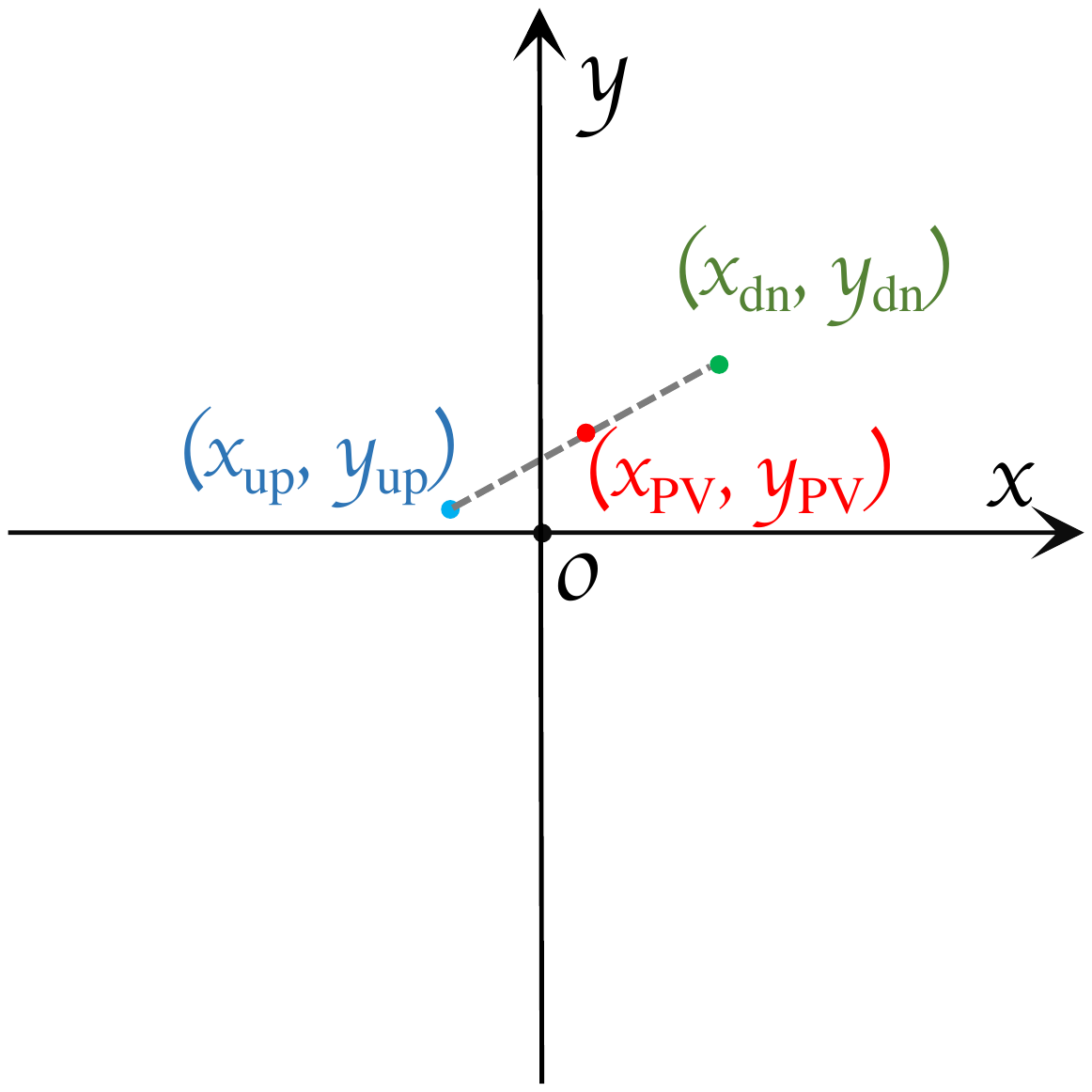}\label{fig:IP_xy}}
    \subfigure[]{\includegraphics[width=0.35\textwidth]{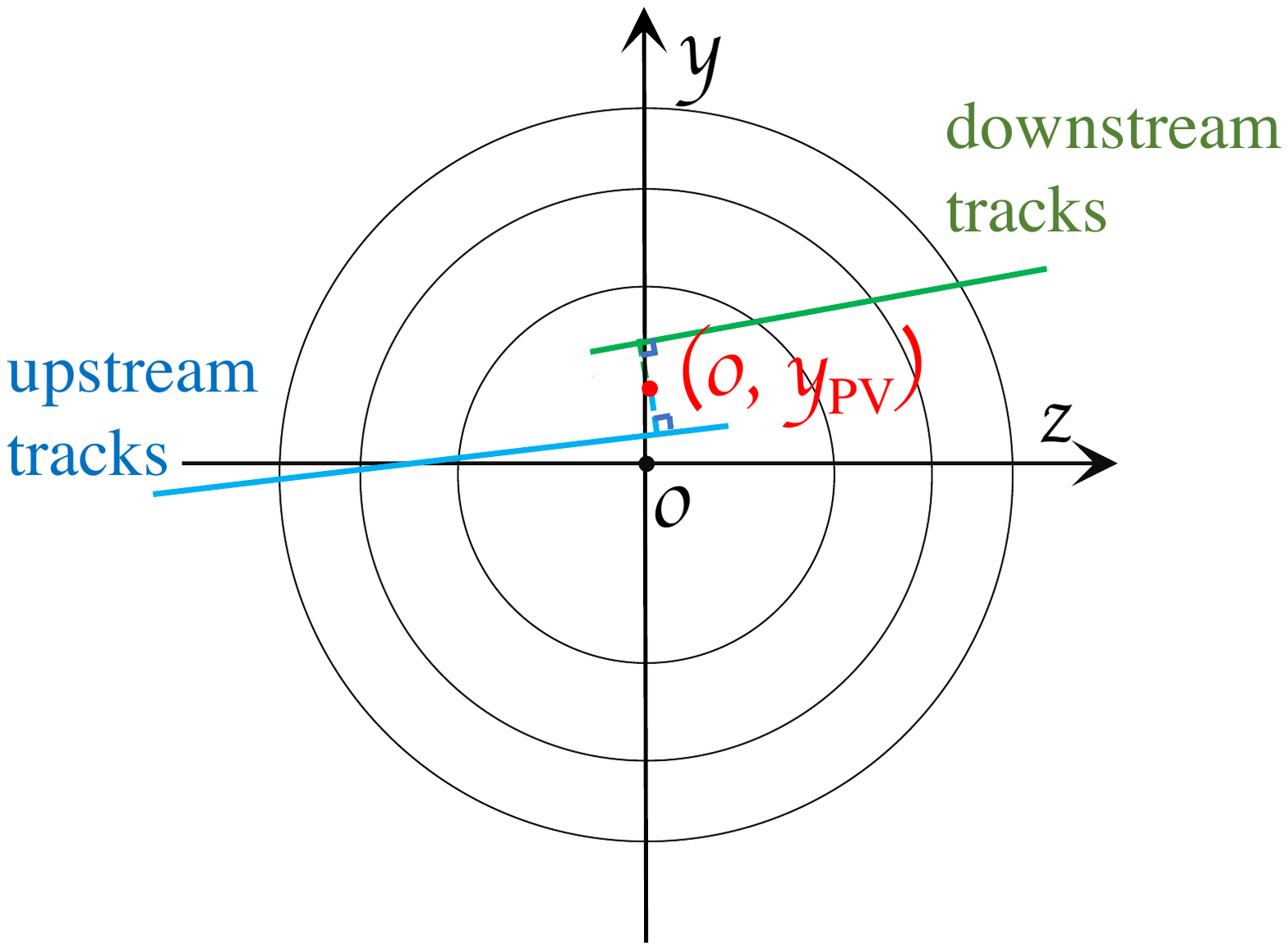}\label{fig:IP_yz}}
    \caption{\subref{fig:IP_xy} $(x_{\rm up}, y_{\rm up})$ and $(x_{\rm dn}, y_{\rm dn})$ represent the extrapolated positions of upstream track and downstream track at $z = 0$ plane. The point $(x_{\rm PV}, y_{\rm PV})$ denotes the midpoint between $(x_{\rm up}, y_{\rm up})$ and $(x_{\rm dn}, y_{\rm dn})$. \subref{fig:IP_yz} Projection to the $y-z$ plane. The impact parameter is the perpendicular distance between the PV and the tracks.}
    \label{fig:IP_axis}
    \end{center}
\end{figure}

The impact parameter is defined as the perpendicular distance between the track and the primary vertex (PV). In the case of this beam test, the electron beam directly passed through six ladders from one side of the vertex detector prototype. Each track is split into an upstream track and a downstream track, based on hit points from the first three ladder layers and the last three ladder layers, respectively. The upstream track and downstream track are fitted separately. A loose $\chi2$ cut for track quality is applied, with $\chi2 / N_{\rm DoF} < 4$. As depicted in Fig.~\ref{fig:IP_axis}\subref{fig:IP_xy}, the primary vertex $(x_{\rm PV}, y_{\rm PV})$ is assumed to be the midpoint between the two points $(x_{\rm up}, y_{\rm up})$ and $(x_{\rm dn}, y_{\rm dn})$, where the upstream and downstream tracks extrapolated to the $z = 0$ plane. In Fig.~\ref{fig:IP_axis}\subref{fig:IP_yz}, the impact parameter is calculated as the perpendicular distance from the primary vertex $(x_{\rm PV}, y_{\rm PV})$ to either the upstream or downstream track. Even though the impact parameter is not strictly well-defined, it can still reflect the overall performance of this vertex detector prototype.

\begin{figure}[ht]
    \begin{center}
    \subfigure[]{\includegraphics[width=0.23\textwidth]{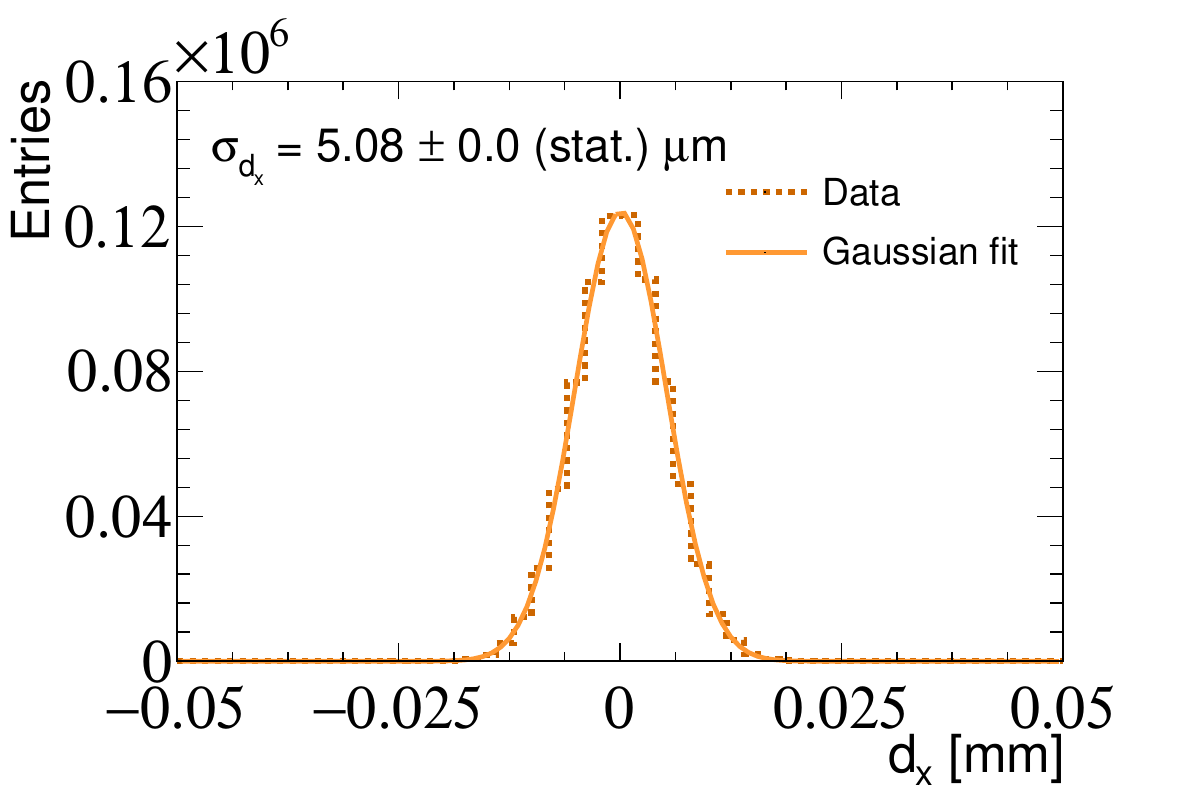}\label{fig:IPres_x}}
    \subfigure[]{\includegraphics[width=0.23\textwidth]{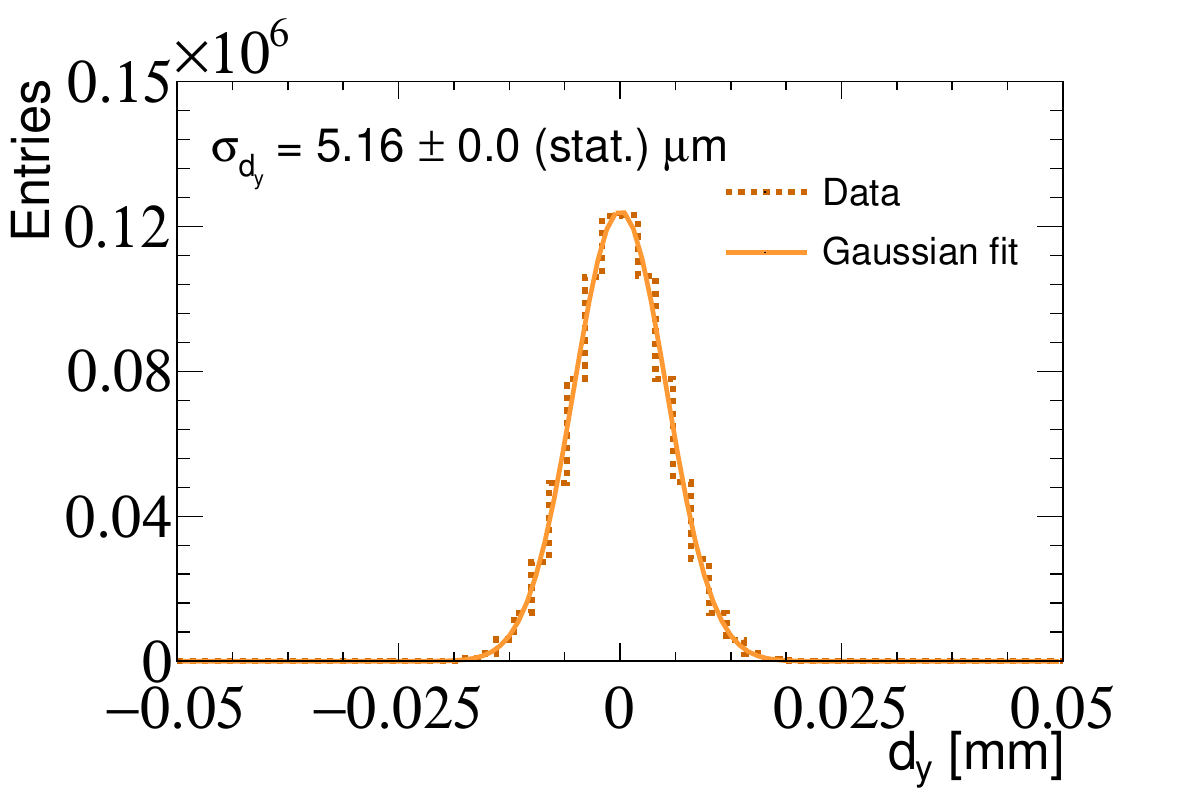}\label{fig:IPres_y}}
    \caption{The distribution of perpendicular distance between upstream tracks and PV in the $x-z$ plane~\subref{fig:IPres_x} and $y-z$ plane~\subref{fig:IPres_y}.}
    \label{fig:IP_res}
    \end{center}
\end{figure}

Therefore, Fig.~\ref{fig:IP_res} shows the perpendicular distances of the primary vertex and the upstream track in $x-z$ plane and $y-z$ plane, with a resolution of $5.08 \pm 0.18$~(syst.)~\SI{}{\um} in the $x$-direction and $5.16 \pm 0.19$~(syst.)~\SI{}{\um} in the $y$-direction for the impact parameter, respectively.

\section{Conclusion}
\label{sec:conclusion}
A first baseline vertex detector prototype developed for CEPC has been tested and characterized using a \SI{6}{\GeV} electron beam at DESY \uppercase\expandafter{\romannumeral2} TB~21. Six ladders with 24 TaichuPix-3 sensors were installed onto the mechanical structure of the prototype. The readout electronics and the DAQ system have been tested and remained stable during the beam test. Analysis of the offline data revealed that the spatial resolution is about \SI{5}{\um} for TaichuPix-3 sensor using Process B in the innermost layer of the prototype, and about \SI{5.4}{\um} for those using Process A in the middle layer of the prototype. The detection efficiency exceedes \SI{99}{\%}. The impact parameter resolution of the prototype, as defined in this article, is about \SI{5.1}{\um}. Reproducing the known performance of the TaichuPix-3 sensors demonstrates the mechanical stability of the setup, as well as the reliability of the electronics readout and DAQ system, indicating the overall good performance of the entire prototype detection system.



\section*{Acknowledgment}

The measurements leading to these results have been performed at the Test Beam Facility at DESY Hamburg (Germany), a member of the Helmholtz Association (HGF). 

\bibliographystyle{IEEEtran}
\bibliography{reference}
\end{document}